\documentclass[preprint]{revtex4}

\usepackage{float}
\usepackage{graphicx}
\usepackage{array}
\usepackage{delarray}
\usepackage{amsmath}
\usepackage{amssymb}
\begin{document}

\newcommand{\myvec}[1]{\accentset{\rightharpoonup}{#1}}
\newcommand{\ket}[1]{| #1 \rangle}
\newcommand{\bra}[1]{\langle #1 |}
\newcommand{\op}[1]{{\mathbf #1}}
\newcommand{\ops}[1]{{\boldsymbol #1}}
\newcommand{\Mit}{\mathrm}
\newcommand{\Tr}[2][]{\mathrm{Tr}_{#1} \! \left[ #2 \right]}

\newcommand{\be}{${}^9\mbox{Be}^+$ \:}
\newcommand{\Sstate}{{}^2{\rm S}_{1/2} \:}
\newcommand{\Pone}{{}^2{\rm P}_{1/2} \:}
\newcommand{\Pthree}{{}^2{\rm P}_{3/2} \:}
\newcommand{\Plev}{\mathrm{P}}
\newcommand{\Fd}{F=2, m_F=-2}
\newcommand{\Fu}{F=1, m_F=-1}
\newcommand{\Fs}{{{\mathcal F}_\mathrm{S}}}
\newcommand{\Fone}{{{\mathcal F}_{1/2}}}
\newcommand{\Fthree}{{{\mathcal F}_{3/2}}}
\newcommand{\Fp}{{{\mathcal F}_\mathrm{P}}}
\newcommand{\Fk}{{{\mathcal F}_k}}
\newcommand{\Eps}{{\hat{\epsilon}}}
\newcommand{\phif}{\phi_\Mit{fluct}}

\newcommand{\ua}{{\uparrow}}
\newcommand{\da}{{\downarrow}}
\newcommand{\uan}{{\uparrow^{\mathrm{N}}}}
\newcommand{\dan}{{\downarrow^{\mathrm{N}}}}
\newcommand{\hc}{\mbox{h.c.}}
\newcommand{\cc}{\mbox{c.c.}}
\newcommand{\deltak}{\myvec{\Delta k}}
\newcommand{\omegaud}{\omega_{\da\ua}}
\newcommand{\omegafs}{\omega_{\rm FS}}
\newcommand{\sumj}{\sum_{j=1,2}}
\newcommand{\sumk}{\sum_{k=\{ 1/2,3/2 \}}}
\newcommand{\Omegaud}{\Omega_{\da\ua}}
\newcommand{\deltaud}{\delta_{\da\ua}}
\newcommand{\Oop}{\boldsymbol{\mathcal{O}}}
\newcommand{\nbar}{\overline{n}_\Mit{COM}}
\newcommand{\heat}{\Gamma_\Mit{heat}}

\newcommand{\mydash}{~$\leftrightarrow$~}

\newcommand{\antih}{$\overline{\mbox{H}}$ }
\newcommand{\htwo}{$\mbox{H}_2$}
\newcommand{\yb}{$\mbox{Yb}^+\:$}

\newcommand{\widtha}{14cm}
\setlength{\abovedisplayskip}{1mm} 

\title{Mesoscopic entanglement of noninteracting qubits using collective spontaneous emission}

\author{D. Kielpinski}

\affiliation{Centre for Quantum Dynamics, School of Biomolecular and Physical Sciences, Griffith University, Brisbane 4111, Queensland, Australia}

\date{\today}
\begin{abstract}

We describe an experimentally straightforward method for preparing an entangled W state of at least 100 qubits. Our repeat-until-success protocol relies on detection of single photons from collective spontaneous emission in free space. Our method allows entanglement preparation in a wide range of qubit implementations that lack entangling qubit-qubit interactions. We give detailed numerical examples for entanglement of neutral atoms in optical lattices and of nitrogen-vacancy centres in diamond. The simplicity of our method should enable preparation of mesoscopic entangled states in a number of physical systems in the near future.

\end{abstract}

\maketitle

The preparation of entangled states is a fundamental task in quantum information processing (QIP). A wide variety of physical systems exhibit long-lived quantum coherence and are potentially useful for QIP, but lack the strong, controllable qubit-qubit interactions that permit entangled state generation \cite{Nielsen-Chuang-QIP-book, Spiller-Kok-QIP-2005-rev}. Various measurement-based quantum computing schemes have demonstrated that the repeated preparation of specific entangled states is an adequate substitute for unitary entangling operations \cite{Gottesman-Chuang-teleportation-QC, Knill-Milburn-linear-optics-QC, Raussendorf-Briegel-one-way-QC}. Here we propose a method for preparing the entangled W state of at least 100 qubits without any direct or indirect qubit-qubit interactions. The W state of $N$ qubits can be written $\ket{W_N} = \ket{100 \ldots 0} + \ket{010 \ldots 0} + \ket{001 \ldots 0} + \ldots + \ket{000 \ldots 1}$. This class of states exhibits unusually robust nonclassical properties \cite{Dur-Cirac-W-state, Sen-Zukowski-W-state-nonclassicality} and can be used for quantum teleportation and secure communication \cite{Joo-Kim-W-state-teleportation, Joo-Kim-W-state-communication}. Our method is applicable to a wide variety of qubit implementations in which the qubit states exhibit spontaneous emission. Detection of a single spontaneous emission photon from a coherently excited assembly of qubits indicates successful state preparation. The highly directional character of the collective emission in free space allows for a robust and efficient detection setup. The simple experimental apparatus required for our method should enable preparation of mesoscopic entangled states in a number of physical systems in the near future. \\

Our method allows straightforward entanglement generation for many attractive QIP candidates that are currently experimentally challenging. Entanglement generation usually requires engineering indirect qubit-qubit interactions through an auxiliary quantum degree of freedom common to all the qubits, a so-called ``quantum bus". The striking success of ion-trap QIP relies on the ion-ion Coulomb repulsion as a naturally occurring quantum bus \cite{Cirac-Zoller-ion-CNOT}. In most cases, though, the quantum bus itself requires considerable effort to construct, and the usual technique is to invoke the highly challenging technology of strong-coupling cavity QED, as in proposals for QIP with neutral atoms \cite{Domokos-Haroche-CQED-CNOT}, with quantum dots \cite{Imamoglu-Small-quantum-dot-CQED-QIP}, and with NV-centers \cite{Shahriar-Craig-NV-center-CQED-QIP}. In contrast, our reliance on detection of collective emission ensures the coherence of the entangled state, even for emission in free space. \\

Collective emission processes have recently found use in generation of single photons \cite{Lukin-light-storage-rev, Chou-Kimble-atomic-ensemble-single-photon} and in the preparation of entanglement between collective excitations of two separate ensembles of atomic qubits \cite{Duan-Zoller-linear-optics-atom-QC, Matsukevich-Kuzmich-spin-wave-entanglement}. The repeat-until-success protocols used in those works are the inspiration for our method, which generates multipartite entanglement of the mesoscopic number of individual qubits making up the ensemble, rather than using a collective excitation as a single qubit. The multipartite entanglement available from our method is more powerful than entanglement of two collective-excitation qubits, and simpler to implement than entanglement of large numbers of collective excitations. Our method also avoids the requirement on collective-excitation protocols to use optically dense qubit ensembles, which can restrict applications to inconveniently high numbers or densities of qubits. \\

\subsection*{Collective spontaneous emission}

Consider an assembly of $N$ two-level atoms with transition wavelength $\lambda$ and decay time $\tau$. The atoms are contained in a region with dimensions much smaller than $c \tau$, but much larger than $\lambda$. The atoms are initially prepared in the ground state and interact with a series of resonant plane-wave laser pulses of wavevector $\vec{k}_L$. The atomic dynamics is conveniently described using the collective operators \cite{Arecchi-Thomas-atomic-coherent-states}

\begin{equation}
J_z = \sum_j \sigma_z^{(j)} \quad J_{+}^{\vec{k}} = \sum_j \sigma_+^{(j)} \exp \left[ i \vec{k} \cdot \vec{x}_j \right]
\end{equation}

\noindent parametrized by wavevectors $\vec{k}$, where the $\sigma^{(j)}$ are the Pauli operators of the $j$th atom. In this formalism, a laser pulse with area $\pi/2$ and phase $\phi_L$ acts as the operator $U_{\pi/2}(\phi_L) = \exp[i(\pi/4) e^{-i \phi_L} J_{+}^{\vec{k}_L} + {\mathrm h.c.}]$. The radiative dynamics is generally fast compared to the motion of the atoms, so we take the atoms to be in eigenstates of the position operator $\vec{x}$.\\

The radiation produced by atomic decay is described by the atom-field interaction Hamiltonian $H_\Mit{int} = \hbar \sum_{\vec{k}} g_{\vec{k}} a_{\vec{k}} J_+(\vec{k}) + \mbox{h.c.}$, where $a_{\vec{k}}$ is the annihilation operator for a plane-wave field along $\vec{k}$. The angular dependence of $g_{\vec{k}}$ gives rise to the single-atom dipole radiation pattern. The intensity distribution of the atomic radiation at short times is found by Fermi's Golden Rule to be \cite{Dicke-spontaneous-emission-coherence}

\begin{align}
I(\vec{k}) &= I_0(\vec{k}) \: \frac{N}{4} (1 + \zeta(\vec{k}))\\
\zeta(\vec{k}) &\equiv \frac{1}{N} | \sum_{j=1}^N \exp (- i (\vec{k} - \vec{k}_L) \cdot \vec{x}_j) |^2
\label{eq:intensity}
\end{align}

\noindent where $I_0(\vec{k})$ is the single-atom radiation pattern. The coherent term $\zeta(\vec{k})$ is approximately equal to $N$ over a narrow range of $\vec{k} \approx \vec{k}_L$, so the radiation pattern contains a narrow peak in the forward direction. The power in the forward peak can be a significant fraction of the total power.\\

\subsection*{Conditional dynamics and entangled-state preparation}

Motivated by this examination, we study the atomic dynamics conditional on emission of a single photon along $\vec{k}_L$. Suppose the atoms are excited with a $\pi/2$ pulse with phase $\phi_L$, radiate freely for a time $T_\Mit{det}$ much shorter than the time needed for emission of a single photon, and are de-excited with another $\pi/2$ pulse with phase $\phi_L + \pi$. If no photon is emitted during $T_\Mit{det}$, the state of the atoms is the same before and after the operations. However, if a photon is emitted along $\vec{k}_L$, the pulse sequence implements the operator \cite{Arecchi-Thomas-atomic-coherent-states}

\begin{align}
M_\Mit{det}(\phi_L) &= U_{\pi/2}(\phi_L+\pi) \: J_-^{\vec{k}_L} \: U_{\pi/2}(\phi_L)\\
&= \frac{1}{2} J_+^{\vec{k}_L} - \frac{1}{2} J_-^{\vec{k}_L} e^{-2 i \phi_L} + \frac{1}{\sqrt{2}} J_z e^{-i \phi_L}
\end{align}

\noindent where we have assumed that the $\pi/2$ pulse durations are always much shorter than $T_\Mit{det}$. Successful single-photon detection during the detection period $T_\Mit{det}$ implies that the atoms are in the state $M_\Mit{det}(\phi_L) \ket{gg\ldots g}$. \\

\begin{figure*}
\begin{center}
\includegraphics*[width=\widtha]{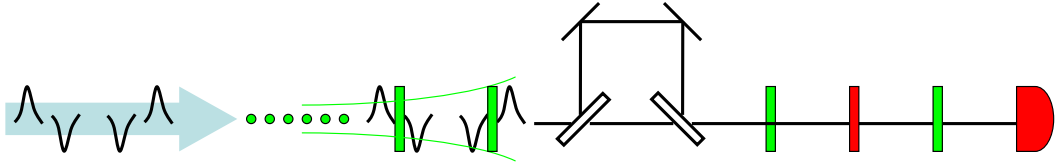}
\caption{Experimental arrangement for preparing the W state of $N$ qubits. Two laser pulse sequences generate small amounts of collective spontaneous emission from the assembly of atoms (green). An unbalanced Mach-Zehnder interferometer and subsequent temporal gating of the overlapped emission (red) allow selective detection. A click at the detector (red) implements the entangling operator $M_\Mit{ent}$.}
\label{expschem}
\end{center}
\end{figure*}

We show the experimental arrangement for entangled state preparation in Fig.~\ref{expschem}. In one trial of the repeat-until-success protocol, we detect the atomic radiation along the forward direction $\vec{k}_L$ during a pair of pulse sequences of the type that implement $M_\Mit{det}$. The laser phase is shifted by $\pi$ between the two pulse sequences. Before detection, the emitted radiation passes through an unbalanced Mach-Zehnder interferometer to induce constructive interference between the two emission time-bins. Detection of a single photon in the interference time-bin implies that the operator

\begin{align}
M_\Mit{ent} &= \frac{1}{2} (M_\Mit{det}(\phi_L) + M_\Mit{det}(\phi_L+\pi))\\
&= \frac{1}{2} (J_+^{\vec{k}_L} - J_-^{\vec{k}_L} e^{-2 i \phi_L})
\end{align}

\noindent has acted on the atomic state. Since $J_-^{\vec{k}}$ annihilates the initial state, we find the final state $J_+^{\vec{k}} \ket{gg \ldots g}$. This is just the first excited Dicke state along $\vec{k}_L$ for an extended atomic ensemble. Regarding each atom as a qubit over the states $\ket{g}$, $\ket{e}$, we see that the final state is related by single-qubit phase factors to the W state of $N$ qubits \cite{Dur-Cirac-W-state}, which we write $\ket{W_N}$. As long as the qubits remain motionless on the scale of a wavelength, we can ignore the additional phases. \\

\subsection*{Sources of error in the entangled state}

As in other repeat-until-success quantum protocols, entanglement purification is built into our scheme by the conditional detection \cite{Duan-Zoller-linear-optics-atom-QC}. We can analyze the entanglement of the final state using the entanglement witness ${\mathcal W} = (1-1/N) \openone - \ket{W_N}\bra{W_N}$ \cite{Bourennane-Sanpera-entanglement-witness, Haffner-Blatt-scalable-entanglement}. For any density matrix $\rho$, the condition ${\rm Tr}[{\mathcal W} \rho] = 1 - 1/N - F < 0$ implies that $\rho$ is entangled. The infidelity $1 - F$ is never greater than the sum of the various error probabilities, so the final state is certainly entangled as long as the sum of the error probabilities is less than $1/N$. \\

For atom number densities less than $1/\lambda^3$, the total emission probability $P_\Mit{emis}$ during each trial considerably exceeds the probability of successful detection $P_\Mit{det}$. However, this inefficiency need not affect the fidelity of state preparation. If a photon is detected, indicating a successful trial, there is a probability $P_\Mit{emis}$ that a second photon has also been emitted during the detection period. The final state resulting from the double-emission event is not the desired entangled state. This error source can be reduced to any desired level by shortening the detection time, at the cost of a lower success probability per trial.\\

A similar error source arises from the uncertainty in the wavevector of the detected photon $\vec{k}_\Mit{det}$. Mismatch between $\vec{k}_\Mit{det}$ and $\vec{k}_L$ induces an additional phase $\phi_j = (\vec{k}_\Mit{det} - \vec{k}_L) \cdot \vec{x}_j$ on the $j$th atom, so that we obtain the state $\ket{W'_N} = [\Pi_j \exp(-i \phi_j \sigma_z^{(j)})] \ket{W_N}$. Averaging the fidelity $|\langle W_N | W'_N \rangle|^2 = \zeta(\vec{k})/N$ over the cone of detected wavevectors, we find the error in the final state  to be $1 - \overline{\zeta}/N$. We can reduce this source of error by restricting the angular range of detection to keep $\zeta(\vec{k}) \sim N$, again at a cost of lower success probability. \\

Finally, successful entangled state preparation relies on the initialization of the atomic state to $\ket{gg \ldots g}$ before each trial. An unsuccessful trial implements the operator $1 + \sum_{\vec{k} \neq \vec{k}_L} \epsilon_{\vec{k}} J_+^{\vec{k}}$, so that each atom is left in the excited state with a small probability. To initialize the state for the next trial, we allow the atoms to decay for a time $T_\Mit{init}$. The total initialization error $\kappa_\Mit{init}$ can then be estimated conservatively as the error in a single trial, $\exp(-T_\Mit{init}/\tau)$, times the number of trials. While the initialization step is by far the slowest step in an individual trial, the exponential scaling of $\kappa_\Mit{init}$ implies that the initialization time per trial scales only as $\log N$ for entanglement of $N$ particles. \\

\subsection*{Implementations: neutral atoms and NV centres}

As an example, we consider the application of our method to a string of neutral atoms placed at adjacent antinodes of an optical lattice. This situation can be realized experimentally in the near future, due to recent advances in micromanipulation of neutral atoms \cite{Schrader-Meschede-neutral-atom-quantum-register, Miroshnychenko-Rauschenbeutel-two-atom-insertion}. Those experiments used Cs atoms in a lattice with 532 nm spacing between antinodes. For circularly polarized laser excitation on the ${}^2{\mathrm S}_{1/2} \ket{F=4,m_F=4} \rightarrow {}^2{\mathrm P}_{3/2} \ket{F=5,m_F=5}$ cycling transition at 852 nm, the atoms behave as two-level systems. Laser excitation pulses of femtosecond duration and energies of a few nJ can implement essentially instantaneous $\pi/2$ rotations, and are readily obtained from mode-locked Ti:sapphire lasers. Temporally gated detection of the atomic emission at the single-photon level can be accomplished by upconverting the emission with an auxiliary laser pulse and spectral filtering to detect only the upconverted signal \cite{Mataloni-DeMartini-upconversion-single-photon-detection, Albota-Wong-telecom-upconversion-detection, Vandevender-Kwiat-single-photon-upconversion}.\\

\begin{figure}
\begin{center}
\includegraphics*{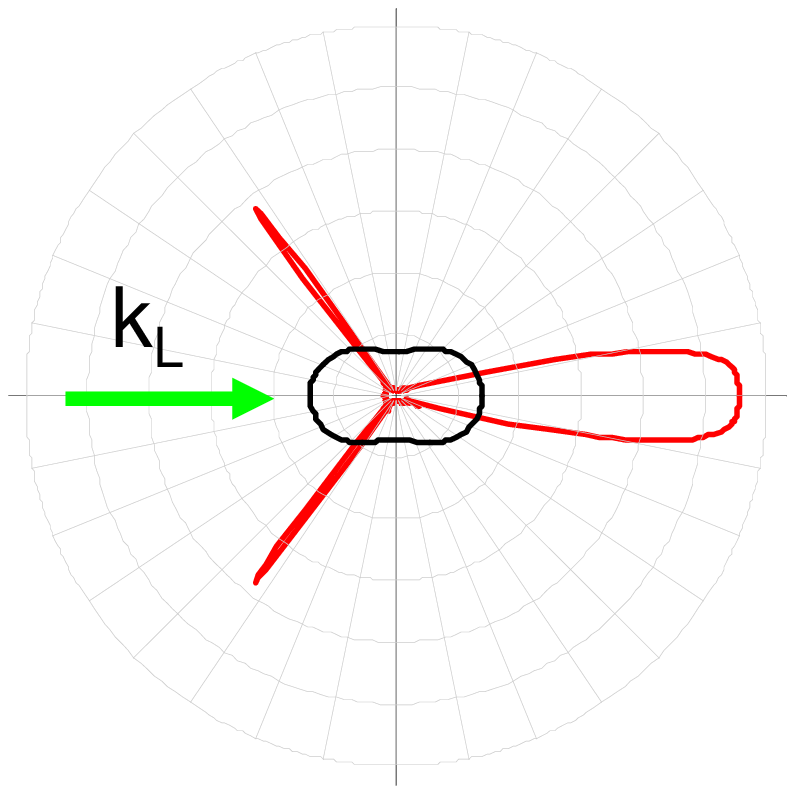}
\caption{Angular intensity distribution of emitted radiation (red) for thirty Cs atoms in an optical lattice. The atoms (not shown), at the center of the plot, are equally spaced by 532 nm along the direction $\vec{k}_L$ (green) of the exciting laser pulse. The intensity distribution is symmetric under rotation about $\vec{k}_L$. The collective forward emission peak is absent in the angular distribution for a single atom (black, not to scale).}
\label{neutralemis}
\end{center}
\end{figure}

Fig.~\ref{neutralemis} shows the emission pattern predicted by Eq.~(\ref{eq:intensity}) for a string of 30 atoms lying parallel to $\vec{k}_L$. The usual single-atom intensity distribution under $\sigma^+$-polarized excitation is $I_0(\vec{k}) \propto (1 + \cos^2\theta)$, where $\theta$ is the angle between $\vec{k}$ and $\vec{k}_L$. Table~\ref{params} gives examples of entangled state preparation with 10, 30, and 100 atoms, with detection parameters chosen so that each error source contributes an error of less than $0.2/N$, ensuring the final state is entangled. We define $S$ as the spontaneous emission rate in units of $1/\tau$. Since $S$ is approximately 1 in all cases, the atoms are not in the superradiant regime. The preparation times of a few ms are on the order of the time required for generation of six- and eight-ion entanglement in recent experiments \cite{Leibfried-Wineland-six-atom-entanglement, Haffner-Blatt-scalable-entanglement}. Notably, the sum of the computed errors in preparing the 100-atom W state is considerably below the current gate error threshold for fault-tolerant quantum computation \cite{Knill-three-percent-error-correction}. \\

\begin{table}
\begin{center}

\caption{Parameters for entangled state preparation with Cs atoms and NV centres. $\tau$, upper-state lifetime. $N$, number of qubits. $S$, spontaneous emission rate per atom in units of $1/\tau$. $\alpha_\Mit{det}$, half-angle of detection cone. $\eta_\Mit{det}$, photon fraction emitted into detection cone. $T_\Mit{det}$, detection time. $T_\Mit{init}$, initialization time. $N_\Mit{tr}$, number of trials to prepare entangled state with 50\% probability. $T_\Mit{prep}$, time required to prepare entangled state with 50\% probability.}
\label{params}

\renewcommand{\arraystretch}{1.5}

\[
\begin{array}{l|c|c|c|c|c|c|}
& \multicolumn{3}{c|}{\mbox{Cs, $\tau$ = 30 ns}} & \multicolumn{3}{c|}{\mbox{NV, $\tau$ = 13 ns}} \\
\hline
N & 10 & 30 & 100 & 10 & 30 & 100 \\
S & 1.2 & 1.2 & 1.2 & 1.9 & 3.0 & 5.7 \\
\alpha_\Mit{det} \mbox{[mrad]} & 210 & 91 & 37 & 280 & 150 & 62 \\
\eta_\Mit{det} & 0.074 & 0.040 & 0.021 & 0.17 & 0.082 & 0.026 \\
T_\Mit{det} \mbox{[ps]} & 51 & 5.6 & 0.50 & 1.6 & 0.96 & 0.046 \\
T_\Mit{init} \mbox{[ns]} & 255 & 330 & 400 & 150 & 160 & 200 \\
N_\Mit{tr} & 470 & 2600 & 1.6 \times 10^4 & 1800 & 1300 & 1.3 \times 10^4 \\
T_\Mit{prep} \mbox{[}\mu{\rm s]} & 120 & 840 & 6400 & 270 & 200 & 2800 \\
\end{array}
\]

\end{center}
\end{table}

Our entanglement preparation method is useful for a broad range of qubit implementations, in particular for ensembles of nitrogen-vacancy (NV) centres in diamond. The NV-centre defect exhibits a simple level structure with a zero-phonon line near 637 nm, and has been intensively studied for QIP applications \cite{Wrachtrup-Jelezko-NV-QIP-rev, Greentree-Prawer-NV-QIP-rev}. Optical pumping experiments with NV-centre ensembles have achieved up to 80\% spin polarization of the ground ${}^3$A state \cite{Harrison-Manson-NV-optical-pumping}, and inhomogeneous linewidths as low as 20 GHz have been observed on the zero-phonon line \cite{Santori-Prawer-NV-center-CPT}. The $\Delta m_S = 0$ selection rule on the zero-phonon line ensures that initially spin-polarized centres behave as near-perfect two-level systems under optical excitation \cite{Manson-Sellars-NV-spectroscopy-rev}, with dipole emission pattern $I_0(\vec{k}) \propto \cos^2\theta$ for $\pi$-polarized excitation. Table~\ref{params} gives parameters for cylindrical ensembles of NV centres, using the intensity distribution derived for a cylinder by \cite{Rehler-Eberly-superradiance} and a value of 13 ns for the single-emitter decay time \cite{Manson-Sellars-NV-spectroscopy-rev}. The volume of the cylinder is chosen to maintain a constant number density of $2 \times 10^{14} \:\mbox{cm}^{-3}$, and the length is taken to be ten times the diameter, yielding superradiant behavior ($S \gg 1$). As before, all detection parameters are chosen so that each error source contributes an error less than $0.2/N$, ensuring the final state is entangled. Unlike the case of Cs atoms, the inhomogeneous broadening induces phase rotations between NV centres, but the accumulated phase is small for sufficiently short detection times. The resulting error is smaller than the other error sources for the examples in Table~\ref{params}, and limits the detection time only in the 10-qubit example. \\

\subsection*{Conclusion}

We have presented a repeat-until-success protocol for entangling noninteracting qubits by detecting collective spontaneous emission. The protocol is applicable to a wide range of qubit implementations that exhibit spontaneous emission. The experimental setup is simple and robust, involving only laser pulses and single-photon detectors, with no need for optical resonators. This simplicity should enable practical realisation of mesoscopic entanglement in the near future. Built-in entanglement purification mechanisms let us reduce important error sources to any desired level by modifying the detection setup. Entangled states of at least 100 qubits can be generated with reasonable experimental parameters and with preparation times comparable to those required for generation of few-ion entangled states \cite{Leibfried-Wineland-six-atom-entanglement, Haffner-Blatt-scalable-entanglement}. The protocol is especially useful for solid-state qubit implementations such as NV centers and quantum dots because of its resistance to inhomogeneous broadening. \\

\subsection*{Acknowledgments}

I thank V. Vuletic, H.M. Wiseman, K.R. Brown, and J. Twamley for helpful discussions. The author was partly supported by the Pappalardo Fellowship Program.

\end{document}